# Frequency Point Game Environment for UAVs via Expert Knowledge and Large Language Model

Jingpu Yang[1], Hang Zhang[1], Fengxian Ji[2], Yufeng Wang[1, *], Mingjie Wang[3, *], Yizhe Luo[4], Wendui Ding[1]

[1]*Beihang University, Beijing, 100191, China*

[2]*Northeastern University, Shenyang, Liaoning Province, 110004, China*

[3]*Academy for Network and Communications of CETC, Shijiazhuang, Hebei Province, 050081, China*

[4]*Zhengzhou University, Zhengzhou, Henan Province, 450001, China*

**Abstract**

Unmanned Aerial Vehicles (UAVs) have made significant advancements in communication stability and security through techniques such as frequency hopping, signal spreading, and adaptive interference suppression. However, challenges remain in modeling spectrum competition, integrating expert knowledge, and predicting opponent behavior. To address these issues, we propose UAV-FPG (Unmanned Aerial Vehicle - Frequency Point Game), a game-theoretic environment model that simulates the dynamic interaction between interference and anti-interference strategies of opponent and ally UAVs in communication frequency bands. The model incorporates a prior expert knowledge base to optimize frequency selection and employs large language models for path planning, simulating a "strong adversary". Experimental results highlight the effectiveness of integrating the expert knowledge base and the large language model, with the latter significantly improving path planning in dynamic scenarios through iterative interactions, outperforming fixed-path strategies. UAV-FPG provides a robust platform for advancing anti-jamming strategies and intelligent decision-making in UAV communication systems.

**Keywords:** Reinforcement learning; Interference decision making; Anti-interference strategy; Expert knowledge base; LLM path planning

## 1. Introduction

In recent years, UAV communication technology has found widespread applications in various fields, such as military [1], agriculture [2], logistics [3], emergency rescue [4], environmental monitoring [5], urban management [6], and power inspection [7]. Using techniques such as frequency hopping [8,9], signal spreading [10,11], and adaptive interference mitigation [12,13], UAVs can maintain stable communication links even in complex electromagnetic environments, allowing them to effectively perform tasks such as reconnaissance, target tracking, and situational awareness. However, as UAVs are increasingly deployed in diverse scenarios, ensuring the security and stability of their communication links is more challenging than ever. Consequently, research on interference and anti-interference mechanisms for UAV communication frequency bands now constitutes a pivotal area in the wireless communications domain [14]. In this context, the competition over frequency resources between opponent parties, coupled with the continuous evolution of interference techniques, has become a core issue that requires substantial attention.

Despite recent progress in UAV anti-jamming techniques [15,16], existing research still exhibits significant deficiencies in simulating adversarial intensity and friendly decision-making capabilities within simulated environments. For instance, the FANETs framework [17] lacks an explicit model for the spectrum game environment, while the application scenario of GPDS [18] is unsuitable for the independent operational needs of UAV swarms in complex electromagnetic environments. These specific limitations point to two deeper challenges: first, a critical research gap exists due to the lack of an effective and systematic model for the complex spectrum confrontation process; second, this modeling deficit directly results in a lack of effective methods for existing anti-jamming strategies to make efficient and adaptive decisions in a dynamic game. Therefore, constructing a simulation and deduction environment that can model this game-theoretic confrontation is especially important and urgent, as it provides a critical pathway for systematically developing and validating anti-jamming decision methods.

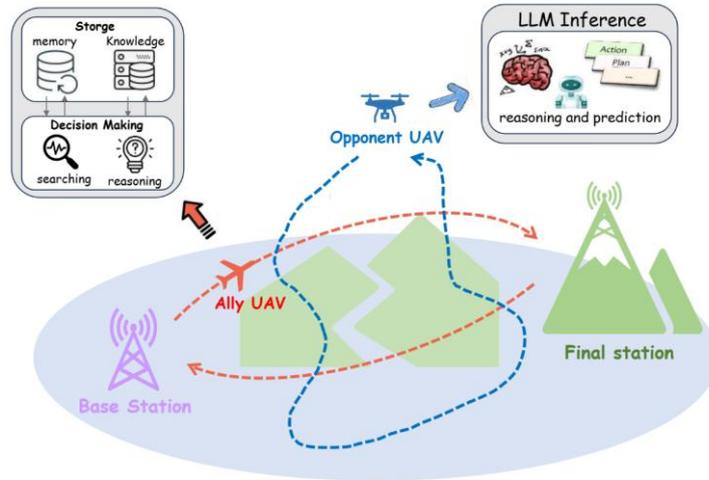

**Fig.1. Illustration of the proposed UAV-FPG environment.** UAV-FPG depicts the UAVs communication system interacting with a base station, ally and opponent UAVs, as well as interactions with an expert knowledge base and large language models to enhance decision-making capabilities.

To address these challenges, we propose a novel environment model named **UAV-FPG**, designed to simulate the dynamic game between interference and anti-interference in UAV communication frequency bands. Our approach unfolds in three key stages. First, we construct a reinforcement learning-based signal game environment [19,20], where the Ally UAV employs spread spectrum and frequency hopping to maintain communication, while opponent entities actively use suppression jamming to disrupt the link, providing a realistic experimental platform (Fig.1). Furthermore, we integrate a prior expert knowledge base into the decision-making process of the Ally UAV. This allows for more rational and efficient frequency selection in response to diverse interference tactics from adversaries [21,22]. Lastly, to create a more realistic and challenging opponent, we introduce a large language model (LLM) to perform gradient-free contextual inference and plan the flight path of the adversary UAV. This simulates a "strong adversary" whose behavior is adaptive and less predictable, thereby enhancing the decision-making validation for Ally UAVs in complex scenarios [23]. The primary contributions of this paper are summarized as follows:

(1) We establish the first reinforcement learning environment for signal game theory in UAV communication bands, modeling the dynamic interplay of suppression jamming and anti-jamming techniques (e.g., frequency hopping) to create a high-fidelity platform for studying wireless communication confrontation strategies.

(2) We incorporate a prior expert knowledge base to guide the Ally UAV's frequency selection, enhancing the rationality of its frequency switching decisions and improving its anti-interference capabilities against diverse jamming tactics.

(3) We introduce LLM for intelligent, gradient-free path planning of the adversary UAV. This creates a challenging "strong adversary" that more closely mirrors the complexity of real-world application scenarios, providing an innovative approach to decision-making in complex games.

**2. Related Work**

2.1 Multi-Agent Game Theory

Multi-Agent Game Theory [24-27], as an essential branch of artificial intelligence and reinforcement learning, studies the interactions and strategy formulation of multiple agents in competitive and cooperative environments, and it achieves significant success in scenarios such as non-cooperative games, cooperative games, and zero-sum games. In recent years, the development of Deep Multi-Agent Reinforcement Learning provides powerful tools for this field [28], enabling agents to perform adaptive strategy optimization in dynamic and uncertain environments. In non-cooperative games, commonly used models include self-play and opponent learning, which allow agents to iteratively learn optimal strategies without cooperation [29]. For example, AlphaGo and AlphaZero use self-play to iteratively optimize strategies, achieving success in complex game scenarios such as Go [30]. In cooperative games, frequently employed models include Centralized Training and Decentralized Execution (CTDE) [31] and game-theoretic strategy optimization methods, such as VDN [32] and QMIX [33]. These models enable agents to form effective cooperative strategies in multi-agent collaborative tasks, enhancing overall efficiency and success rates in scenarios like robotic cooperation [34] and multi-UAV control [35].

In zero-sum games, models often rely on opponent modeling and predictive strategies, such as Fictitious Play [36] and Counterfactual Regret Minimization (CFR) [37], allowing agents to dynamically adjust strategies in opponent environments. These models achieve successful applications across multiple fields. In autonomous driving [38], multi-agent game theory is used to address vehicle cooperation and collision avoidance, allowing the system to adapt well in dynamic traffic scenarios. In robotic collaboration [39], multi-agent game theory enables multiple robots to allocate tasks effectively in warehouse management and rescue missions, improving efficiency and reducing conflicts. In financial markets [40], game-theoretic models simulate the behavioral interactions of different traders, helping researchers optimize trading strategies and gain a better understanding of market fluctuations. In military simulations [1], multi-agent games are employed to evaluate and optimize tactical strategies, enhancing the accuracy and effectiveness of decision-making.

In the current UAV domain, multi-agent game theory is widely applied to various aspects such as task allocation, path planning, and cooperative control. However, research related to signal games remains relatively limited [41-43]. To address this gap, we create a new environment for UAV signal games, aiming to explore signal optimization strategies under multi-agent collaboration and competition by simulating different signal game scenarios, which optimizes jamming and anti-jamming strategies between opponent parties through a non-cooperative game mechanism. This environment effectively enhances the adaptability and robustness of base stations when they face opponent signal interference, and it provides strong support for communication security in the UAV domain.

2.2 Incorporation of Expert Knowledge Bases

The integration of expert knowledge bases emerges as a crucial strategy for enhancing the performance and interpretability of intelligent systems, especially in complex tasks that require domain-specific knowledge support. Expert knowledge bases provide structured and domain-relevant information that effectively guides model behavior, improves learning efficiency, and supplies prior support in intricate decision-making environments. This integration achieves significant advancements across various fields, including natural language processing (NLP), healthcare [44-46], and robotics [47], enabling models to better align with established knowledge frameworks and expert rules.

In the realm of NLP, expert knowledge bases [48] such as ontologies and specialized terminological dictionaries play an essential role in enhancing tasks like named entity recognition, sentiment analysis, and question answering. By incorporating domain-specific prior knowledge, for instance, medical terminologies from the Unified Medical Language System (UMLS) [49,50], models attain higher accuracy and improved contextual understanding. Similarly, legal knowledge bases offer robust support for NLP models in processing legal documents, text classification, and information retrieval, which allows for more precise comprehension of specialized legal terminology and contexts [51]. In robotics, particularly within collaborative and autonomous systems, expert knowledge bases provide substantial support in areas such as path planning [52], task execution, [53] and environmental interaction [54]. By integrating knowledge related to target identification, navigation constraints, and safety protocols, robotic systems operate more robustly and efficiently in dynamic environments, thereby significantly enhancing task completion efficiency and safety.

Building on these successful applications, we design a frequency selection model for UAV signal games that incorporates an opponent knowledge base of opponent drones under various interference types and frequencies.

This knowledge base offers prior support for our UAVs' center frequency selection during frequency hopping, enabling rapid adjustments based on frequency management and interference avoidance strategies when faced with hostile signal interference. By embedding these expert strategies into the model, our UAVs swiftly adapt to opponent disruptions, effectively ensuring communication security and stability in complex opponent environments. This approach provides robust support for UAV communication systems under challenging conditions, fully demonstrating the significant value of prior knowledge in practical applications.

2.3 Path Planning with Large Language Models

Path planning, a core problem in unmanned aerial systems, increasingly benefits from the assistance of Large Language Models in recent years [55,56]. Leveraging their powerful capabilities for knowledge integration and reasoning, these models find widespread applications in various navigation and path planning tasks. Models such as GPT-4 [57], Vicuna [58], PaLM 2 [59], and LLaMA [60] are capable of interpreting environmental descriptions, states, and rewards to generate suitable path planning suggestions. In traditional path planning, the introduction of LLMs significantly enhances flexibility and adaptability, particularly in high-dimensional, dynamically changing complex environments where conventional methods struggle [61].

Recent studies, such as WayPoint, explore the potential of LLMs in generating goal-directed paths by combining natural language with visual information to create feasible navigation routes [62-64]. Similarly, multimodal models like LLaVA utilize both image and textual information to formulate more precise navigation strategies [65,66]. These studies indicate that LLMs possess considerable potential in complex decision-making tasks, progressively optimizing strategies through interaction with the environment and consideration of constraints. Moreover, certain projects combine LLMs with reinforcement learning to improve path planning performance. For instance, DeepMind's Gato model demonstrates outstanding performance in multitask navigation and control [67], while FLAN-T5 [68,69], through instruction fine-tuning, shows consistent performance in cross-task path planning.

In our work, we employ LLMs for UAV path planning, aiming to enhance strategic behavior in multi-agent games. Specifically, we input the positions and rewards of opponent UAVs from the previous round into an LLM to infer and plan the movement directions of opponent UAVs in the next round, effectively simulating a "strong adversary" effect. This approach effectively leverages LLMs for flight path planning, enhancing strategic flexibility in competitive scenarios and validating the adaptability and learning capabilities of LLMs in addressing complex, dynamic problems in multi-agent games. Path planning based on position and reward offers new avenues for UAV autonomous decision-making, further validating the effectiveness of LLMs in high-dimensional decision-making tasks.

**3. Environment Model**

In the UAV-FPG communication game environment, we focus on signal attenuation between the base station and our UAV, as well as interference caused by opponent UAVs. We simulate interference power attenuation on

UAVs and the intensities of various interference types. Our UAV operates within a designated frequency band, while opponent UAVs detect our communication center frequency and employ power suppression interference to degrade the signal-to-noise ratio (SNR), thereby reducing communication quality. opponent UAVs adopt diverse interference strategies, including single-tone interference, narrowband targeting, wideband jamming, and comb-spectrum interference. Utilizing reinforcement learning agents, opponent UAVs select interference actions, adjust their center frequencies, and optimize their strategies based on reward signals.

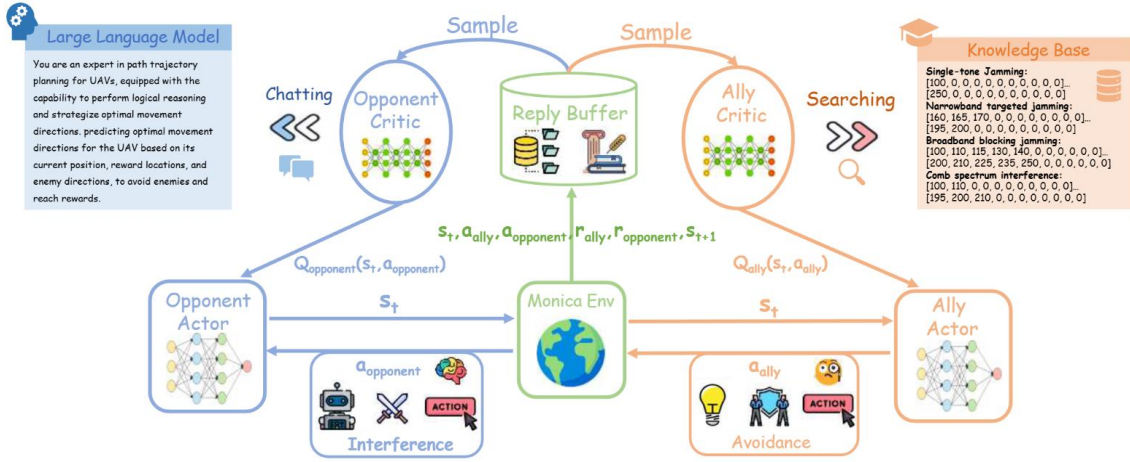

**Fig.2. Framework of the multi-agent reinforcement learning-based UAV-FPG environment.** The game environment employs an actor-critic model architecture. To optimize learning and strategic decision-making, the system integrates anti-jamming data from an expert knowledge base and path planning from an LLM during the game between opponent and ally drones. Interaction with the UAV-FPG environment allows for the collection of experience: at each timestep, the state $s_t$ encapsulates key information. Based on this state, the ally actor selects an avoidance action, $a_{ally}$, while the opponent actor selects an interference action, $a_{opponent}$. The complete interaction tuple, ($s_t$, $a_{ally}$, $a_{opponent}$, $r_{ally}$, $r_{opponent}$, $s_{t+1}$), is then recorded in a replay buffer, which enables iterative improvements in decision-making across multiple episodes.

Our communication system consists of a base station with constant power and a fixed position, while our UAV follows a predetermined path. To mitigate interference, the UAV employs spread spectrum and frequency hopping techniques. By leveraging an expert knowledge base, the UAV selects interference-free frequencies during hopping and despreading processes, thereby enhancing its anti-interference capabilities. To further increase the complexity of the scenario, we integrate an LLM to infer and plan opponent UAV interference trajectories, thereby maximizing interference against our UAV. As illustrated in Fig.2, this setup ensures realistic and challenging opponent conditions, facilitating the evaluation and improvement of our UAV's resilience and communication performance under interference.

## 4. Methods

In wireless communication systems, suppressive power interference is a persistent challenge, with the power of interfering signals closely related to the distance between the opponent and ally UAV. To evaluate whether the opponent UAV's signal interferes with the ally's communication, we analyze whether the interference frequency range overlaps with the ally's center communication frequency. This chapter details the methods used to construct a signal interference and anti-interference game between the ally and opponent UAVs. In Section 4.1, we introduce a fundamental frequency point game model based on modern mobile communication principles. Section 4.2 explains how prior expert knowledge on the types of interference and interference frequencies is incorporated into the model for selecting the ally UAV's communication frequency. In Section 4.3, we explore and validate the potential of large language models for path inference and planning, as well as introduce prompt engineering techniques for effective path planning, thus creating the ``strong opponent'' effect.

Table 1 The interference intensity coefficient k corresponding to different jamming techniques employed by the opponent [70].

| Jamming Type | k |
| --- | --- |
| Single-tone Jamming | 1.5 |
| Narrowband Targeted Jamming | 1.2 |
| Broadband Blocking Jamming | 0.4 |
| Comb Spectrum Jamming | 0.8 |

4.1 Damage analysis of the target

In the UAV-FPG environment, we construct a frequency game scenario involving ally and opponent UAVs to simulate the frequency interference and avoidance strategies between our side and the adversary in wireless communication. This environment consists of a base station and two UAVs (ally and opponent). The base station selects 15 frequency points within the range of 150 MHz to 250 MHz. The ally UAV, acting as the interference-avoidance agent, primarily employs spread spectrum and frequency hopping techniques to mitigate interference from the opponent UAV and maintain communication with the base station. Meanwhile, the opponent UAV is responsible for implementing suppressive signal interference, which includes techniques such as single-tone interference, narrowband targeted interference, broadband jamming, and comb spectrum interference, in an attempt to obstruct the communication between the ally UAV and the base station. Additionally, we employ the free-space path loss model to account for the energy attenuation of electromagnetic waves as they propagate through the air (see Eq. 1).

$$L_{\text{loss}} = 32.44 + 20 \log d(\text{km}) + 20 \log f(\text{MHz}), \quad (1)$$

where d denote the distance between the base station and ally UAV, and let f represent the center frequency of ally UAV's communication. Considering the attenuation from the base station and the opponent UAV, the received power I of the ally UAV under opponent interference is:

$$I(\text{dBm}) = P_{base} - L_{base} - (k \cdot P_{opponent} - L_{opponent}), \quad (2)$$

where $P_{base}$ represents the transmission power from the base station, $L_{base}$ denotes the free-space path loss power from the base station to the ally UAV, k is the interference intensity coefficient corresponding to different interference methods of the opponent (see Table 1), $P_{opponent}$ represents the interference power transmitted by the opponent UAV, and $L_{opponent}$ denotes the free-space path loss power from the opponent UAV to the ally UAV. Therefore, the channel capacity C received by the ally UAV is:

$$C = B(\text{Hz}) \cdot \log_2 \left( 1 + \frac{10^{\frac{I(\text{dBm})-30}{10}}}{10^{\frac{N_0(\text{dBm})-30}{10}} + 10^{\frac{N_i(\text{dBm})-30}{10}}} \right), \quad (3)$$

where B represents the channel bandwidth, $N_0$ denotes the power of natural Gaussian white noise, and $N_i = k * P_{opponent} - L_{opponent}$, represents the interference power experienced by the ally UAV from the opponent UAV. In the case of spread spectrum, the increase in bandwidth results in a decrease in the SNR of the ally UAV, thereby making it challenging for the opponent UAV to effectively observe the center frequency of the ally UAV. In the case of spread spectrum, the increased bandwidth reduces the SNR of the ally UAV, making it difficult for the opponent UAV to effectively observe its center frequency. To mitigate this, we designed a spreading cost function, as presented in Equation 4, encouraging the ally UAV to de-spread early when necessary, while discouraging frequency hopping due to its high cost H.

$$D(t) = \begin{cases} 0.5t, & \text{if } t \leq 10, \\ 5 + 1.0(t-10), & \text{if } 10 < t \leq 20, \\ 15 + 1.2(t-20), & \text{if } 20 < t \leq 40, \\ 39 + 1.5(t-40), & \text{if } t > 40. \end{cases} \quad (4)$$

The opponent UAV attempts to detect the ally's center frequency every n seconds but fails if the ally's SNR falls below a threshold M. Detection resumes after de-spreading and SNR recovery. The ally UAV follows a fixed path, while the opponent UAV can either patrol along a preset route or plan its movements based on the previous round's position and reward using a large model. Considering factors such as SNR variations, the distance between UAVs, and action costs, we derived the reward functions for both ally and opponent UAVs (see Eq. 5 and Eq. 6).

$$r_{\text{ally}} = \text{SNR} - M - D(t) - H \cdot \begin{cases} 1, & \text{if } a_{\text{hopping}} > 0.5 \\ 0, & \text{if } a_{\text{hopping}} \leq 0.5 \end{cases}, \quad (5)$$

$$r_{\text{opponent}} = E \cdot \theta(30 - \|\mathbf{p}_{\text{opponent}} - \mathbf{p}_{\text{ally}}\|) + \alpha \cdot \Delta\text{SNR}, \quad (6)$$

where $a_{hopping}$ is a parameter trained by a reinforcement learning network, belonging to the interval [0, 1], E represent the distance reward coefficient. $p_{opponent}$ and $p_{ally}$ represent the spatial position vectors of the opponent UAV and the ally UAV. The term △SNR denotes the change in SNR, which quantifies variations in the quality of the communication signal under different conditions. The notation |·| indicates the Euclidean distance, employed to evaluate the spatial separation between the opponent and ally UAV. The function θ(x) is the unit step function, which outputs 1 when a specified threshold condition is met and 0 otherwise. Furthermore, the coefficient α quantifies the reduction in our communication SNR attributable to the presence of opponent UAVs, effectively integrating the impact of opponent interference into the model.

Thus, the entire game process is characterized by the opponent UAV continuously approaching the ally UAV to impose interference, while the ally UAV employs spread spectrum techniques and reselects communication frequencies to evade the interference. Once the opponent UAV detects the ally's central frequency again, it resumes its approach and interference, thereby establishing a persistent frequency-based game. The detailed pseudo-code for this environment is presented in Algorithm 1. The overall time complexity of the algorithm can be simplified and estimated as O(N*P), where N denotes the number of training steps and P represents the parameter scale of the neural networks.

**Algorithm 1 UAV-FPG**
1: **Require:** Initialize actor networks $\pi_{\phi_{ally}}$, $\pi_{\phi_{opponent}}$ and critic networks $Q_{\theta_{ally}}$, $Q_{\theta_{opponent}}$ with random parameters $\theta_{ally}, \theta_{opponent}, \phi_{ally}, \phi_{opponent}$;
2: **Require:** prompt $P$, opponent locations $LLM_{opponent}\{\}$;
3: **Require:** Initialize environment state $s$, replay buffer $\mathcal{B} \leftarrow \emptyset$;
4: **for** $t = 1$ to $N$ **do**
5:     Ally selects action $a_{ally}$ using $\pi_{\phi_{ally}}$;
    $a_{ally} = \pi_{\phi_{ally}}(s_t) + \epsilon$,    $\epsilon \sim \mathcal{N}(0, \sigma^2)$;
6:     Obtain the ally center frequency $f_{ally}$ using an expert knowledge base.
7:     Opponent selects action $a_{opponent}$ using $\pi_{\phi_{opponent}}$;
    $a_{opponent} = \pi_{\phi_{opponent}}(s_t) + \epsilon$,    $\epsilon \sim \mathcal{N}(0, \sigma^2)$;
8:     Obtain new state $s'$ and reward $r_{opponent}, r_{ally}$;
**Environment Step:**
9:     Store transition $(s, a_{ally}, a_{opponent}, r_{ally}, r_{opponent}, s')$ in $\mathcal{B}$;
10:    Store the set of opponent positions in $LLM_{opponent}\{\}$;
11:    **if** episode ends **then**
12:       Reset the environment and state $s$;
13:       Update trajectory using LLM;
14:    **end if**
15: Sample mini-batch: $(s, a_{ally}, a_{opponent}, r_{ally}, r_{opponent}, s') \sim \mathcal{B}$;
16:    **Critic Update:**
      Compute $Q$ targets and update $Q_{ally}, Q_{opponent}$;
17:    **Actor Update:**
      Update $\pi_{ally}, \pi_{opponent}$ by maximizing $Q$ values;
18:    Visualize ally and opponent UAV trajectories;
19: **end for**

4.2 Optimized Frequency Selection with Expert Knowledge

UAVs employ frequency hopping and spread spectrum techniques to counter interference with limited adaptability. However, they struggle to optimize frequency selection under complex interference. This research integrates expert knowledge (e.g., interference strategies, central frequencies) and empirical criteria into datasets to improve decision-making, enhancing communication stability and anti-jamming capabilities. To address this issue, the study analyzes four major types of interference and their corresponding avoidance methods. These interference types include Single-tone Jamming, Narrowband Targeted Jamming, Broadband Blocking Jamming, and Comb Spectrum Jamming, each affecting different frequency ranges and requiring distinct frequency selection strategies to mitigate their impact. Through the examination of these interference types, effective anti-jamming solutions for UAVs in complex electromagnetic environments are developed.

**Single-tone Jamming** is a type of interference that focuses on specific frequency points by imposing a strong jamming signal at a particular frequency, thereby disrupting communication. The jamming model can be expressed as:

$$J_{\text{single-tone}}(f) = A \cdot \delta(f - f_j). \quad (7)$$

To address this issue, we adopt a fast frequency-hopping strategy to avoid interference at the center frequency. Real-time spectrum monitoring is employed to identify the interference frequency $f_{opponent}$, enabling dynamic selection of frequencies that are free from interference.

**Narrowband Targeted Jamming** is a form of interference that operates within a narrow frequency band, specifically designed to target the primary frequency range of a desired signal in order to maximize its disruptive effect. Compared to single-tone jamming, this approach covers a broader range of frequencies while remaining concentrated within a specific band. The mathematical representation of narrowband targeted jamming can be expressed as:

$$J_{Narrow}(f) = \begin{cases} A, & \text{if } f \in [f_c - B/2, f_c + B/2] \\ 0, & \text{otherwise} \end{cases}. \quad (8)$$

To address this, we opt to avoid the interference-affected range by employing a pseudo-random frequency hopping technique, wherein the transmission frequency rapidly switches across multiple channels. This approach effectively evades the interference band, making it difficult for narrowband jamming to continuously disrupt signal transmission.

**Broadband Blocking Jamming** is a type of jamming that spans a wide frequency band, aiming to degrade the SNR of the communication system by injecting noise or interference signals across a broad frequency range. This, in turn, disrupts the receiver's ability to decode the target signal. Unlike narrowband interference, broadband blocking interference is not confined to specific frequency bands but instead affects the entire or a significant portion of the communication spectrum, making it highly disruptive. The expression for this type of interference is given as:

$$J_{\text{Broadband Blocking}}(f) = \begin{cases} P_J, & \text{if } f \in [f_{\min}, f_{\max}] \\ 0, & \text{otherwise} \end{cases}. \quad (9)$$

To address this, we opt to implement frequency hopping to channels outside the interference range, thereby maximizing the distance from the interference band. Alternatively, the original signal's energy can be spread across a wider frequency band, effectively dispersing the impact of broadband interference by averaging its effects over a broader spectrum.

**Comb Spectrum Jamming** is a specialized form of jamming characterized by a spectrum with a comb-like distribution, where single-tone interference signals are injected at specific intervals of frequency points while maintaining no interference at other frequencies. This type of interference selectively occupies certain frequency points within the target communication band, causing significant disruption to frequency-selective signals such as those in OFDM systems. Its mathematical representation can be expressed as:

$$J_{\text{Comb}}(f) = \begin{cases} P_J, & \text{if } f = f_0 + n\Delta f, \, n \in \mathbb{Z} \\ 0, & \text{otherwise} \end{cases} \quad (10)$$

To address this issue, we employ pseudo-random sequences to control frequency hopping, dynamically shifting the operating frequency either to the sensing band or away from the interference region. This approach ensures that the distribution of communication signal frequencies is effectively separated from the interference frequencies.

We have constructed an anti-jamming expert knowledge base incorporating identified interference strategies, center frequencies, and corresponding counter-strategies. By training with this knowledge base, our UAV learn the appropriate anti-jamming techniques, thereby enabling them to effectively counter adversarial interference.

4.3 Strong Opponent Effect: Path Inference and Planning with LLMs

This section presents a path inference and planning approach based on a large language model API, aimed at enhancing the intelligent decision-making capabilities of opponent UAVs in game environments and constructing more challenging opponent scenarios. By integrating reinforcement learning with LLM-based inference mechanisms, we develop a path planning framework that leverages environmental variables as input, enabling opponent UAVs to optimize their action strategies even in complex communication interference environments, such as $f_{\text{opponent}}$.

A systematically designed prompt is formulated to guide the LLM in generating task-specific future movement paths. The prompt incorporates historical positions of the opponent UAV along with their associated rewards, current positional information, and a standardized output format. This ensures that the generated path data is of high quality and accuracy, significantly improving the feasibility and reliability of path inference. Specifically, the core elements of the prompt design include the following:

(1) Historical Positions and Rewards: To help the model understand the behavioral performance of the opponent UAV in past environmental states, each historical position and its corresponding reward are explicitly linked. The expected format is as follows:

"Positions and Rewards: [$x_i$, $y_i$, $z_i$]: $r_{\text{opponent}}$"

(2) Current Position Information: The real-time state of the UAV is conveyed to the model to facilitate the generation of rational paths based on its current status to intercept the ally UAV. The expected format is:

"Current Position: [x, y, z]"

(3) Output Format: A formatted example is provided to clearly define the structure and sequence length constraints for the generated path data, ensuring its usability and structural consistency. The expected format is:

"Next directions: [[$x_1, y_1, z_1$], [$x_2, y_2, z_2$], ..., [$x_n, y_n, z_n$]]"

## 5. Experiments

The present chapter provides a comprehensive description of the experimental design and analysis conducted in the UAV dynamic interaction simulation environment. Firstly, Section 5.1 details the reward-based experiments, wherein the dynamic behavior of allied and opponent UAVs in complex operational scenarios and under opponent conditions is established to evaluate the decision-making capabilities and operational efficiency of the UAVs. Subsequently, Section 5.2 introduces experiments involving path planning using a large language model, exploring the impact of fixed versus dynamic path strategies on enhancing the interference against opponent UAVs. Following this, Section 5.3 presents a dynamic game-theoretic analysis of frequency selection, revealing the evolution characteristics of frequency selection by both allied and opponent UAVs at various stages and demonstrating a significant improvement in the anti-jamming capability of the allied UAV in the latter stages of the game. Finally, Section 5.4 presents an ablation study, separately analyzing the individual contributions of interference methods, expert knowledge bases, and the large language model, thereby assessing their practical impact on decision-making and path planning during opponent engagements. The research presented in this chapter provides significant theoretical support and experimental evidence for the study of autonomous UAV decision-making and opponent strategies.

### 5.1 Environment Setting

This experiment establishes an efficient UAV dynamic interaction simulation environment to investigate the behavioral dynamics of ally and opponent UAVs under various complex operational scenarios and opponent conditions. The experiment leverages RTX 4090 GPU for high-performance computation and utilizes the iFLYTEK Spark Max-32K model API for opponent UAV trajectory planning and decision-making inference. The simulation environment spans a three-dimensional operational space of 1500 * 1500 * 600 meters, where the ally UAV follows a predefined Bezier curve trajectory for round-trip missions, while the opponent UAV operates under four distinct movement patterns[71]: triangle, circle, rectangle, and AI-predicted adaptive trajectories. The environment incorporates Gaussian noise and dynamic interference strategies to simulate realistic electromagnetic interference conditions, Table 2 summarizes the environmental parameters of UAV-FPG. Real-time environmental parameters, such as channel capacity and SNR, are employed to dynamically adjust the ally UAV's strategies, including spreading, frequency hopping, and velocity control, while the opponent UAV adapts its jamming strategies based on opponent dynamics. Table 3 summarizes the hyperparameter values of the Multi-Agent Deep Deterministic Policy Gradient (MADDPG) model. This experimental design integrates real-time interaction, action-reward mechanisms, and complex path planning, aiming to comprehensively evaluate UAV decision-making capabilities, adaptability, and operational efficiency in highly dynamic opponent environments.

Table 2 Parameters of the UAV-FPG game environment.

| Parameter | Value |
| --- | --- |
| Ally bandwidth (non-spread spectrum) | 5 MHz |
| Ally bandwidth (spread spectrum) | 2400 MHz |
| Noise spectral density | -170 dBm |
| Opponent check interval | 5 second |
| UAV speed range | 10 m/s |
| Base station power, $P_{base}$ | 45 dBm |
| Opponent power, $P_{opponent}$ | 20 dBm |
| SNR threshold, $M$ | 8 |

Table 3 Hyperparameters for the MADDPG Model.

| Parameter | Value |
| --- | --- |
| Discount factor, $\gamma$ | 0.99 |
| Total time, $T$ | $10^7$ |
| Batch size, $|B|$ | 32 |
| Learning rate, $\varepsilon$ | 0.001 |
| Buffer capacity, $C$ | $10^6$ |
| State dimension | 15 |
| Action dimension | 9 |
| Max action value | 5 |
| Loss function | MSE |

5.2 Opponent Gameplay Performance in UAV-FPG Environment

We employ large language models for the path planning of opponent UAVs, utilizing two distinct strategies: fixed-path gameplay and dynamic-path gameplay. The fixed-path strategy involves predefining a set of regular patterns, such as triangular, circular, or rectangular trajectories, to evaluate the path planning performance of the LLM. In contrast, the dynamic-path strategy leverages the opponent UAV's position and reward information from each round to generate new trajectories in real time using the LLM. This approach enhances the uniqueness and adaptability of the paths, effectively mitigating trajectory repetition and improving the opponent UAV's interference effectiveness.

Fig.3 illustrates the performance of both static and dynamic path planning strategies executed by opponent UAVs using a LLM in game-theoretic scenarios. The results indicate that both strategies effectively engage in gameplay with allied UAVs, particularly highlighting the superior performance of dynamic path planning. The study demonstrates that multi-round interactive dialogues with feedback using LLM significantly enhance its capabilities in path planning compared to single-round dialogue-based planning.

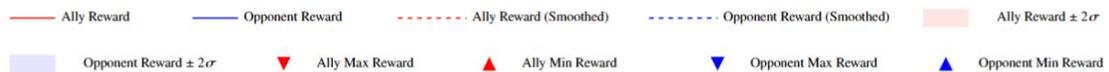

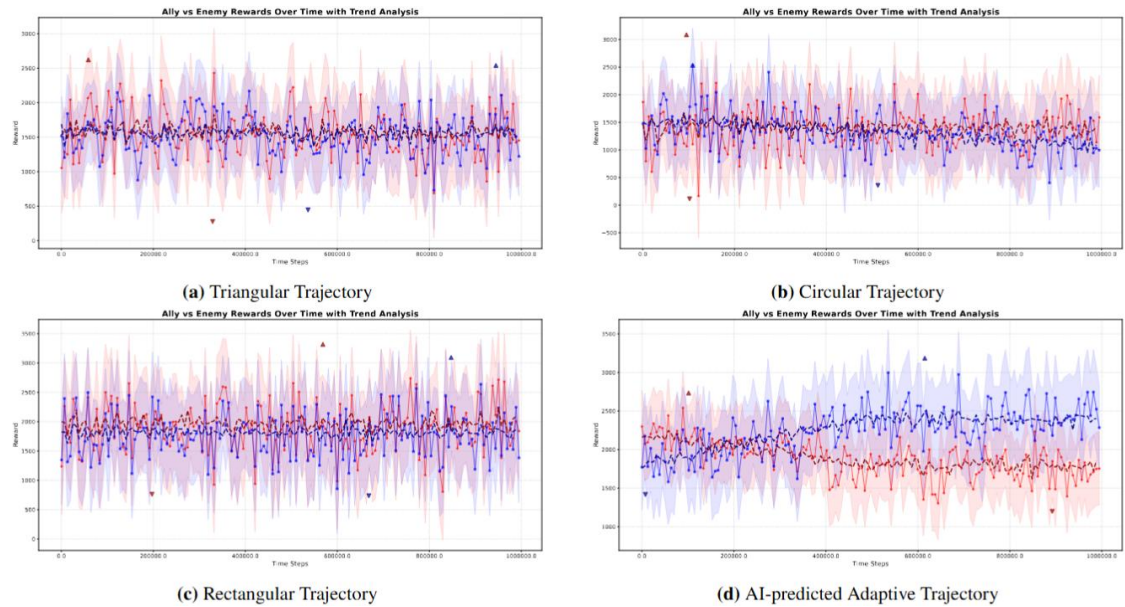

**Fig.3 Comparative and Trend Analysis of Rewards for Ally and Adversary Agents in a Time Series.** The red and blue curves represent the trends of ally and adversary rewards, respectively, with the smoothed curves (dashed lines) further capturing the variations in reward trends. The shaded areas in the figure indicate the standard deviation range (±2σ) of the reward values.

5.3 Analysis of Frequency Selection in Opponent Scenarios

In an opponent communication scenario, we conduct dynamic detection of the communication center frequencies of both the ally and opponent UAVs at different stages, aiming to analyze the dynamic game characteristics of frequency selection comprehensively. Specifically, we monitor the center frequencies of both sides during the initial, middle, and final stages of the game to evaluate the ally UAV's ability to evade opponent signal detection, as well as the opponent UAV's effectiveness in capturing and interfering with the ally's center frequency. This analysis provides theoretical support for understanding the frequency evolution characteristics in opponent communication environments.

In the different stages of the game illustrated in Fig.4, the distribution of the central frequencies between the opponent and ally parties exhibits significant variation. Notably, in the late stages of the game, the overlap between the ally communication central frequencies and the opponent interference frequencies (53.69%) decreases substantially compared to the early (54.46%) and middle (54.39%) stages. Both the overlap ratio and the maximum overlap value show a marked reduction. This indicates that, as the game progresses, particularly in its later stages, the probability of the adversary successfully detecting the allied communication central frequencies diminishes. Consequently, the allied party achieves more effective anti-interference capability in the selection of communication frequencies.

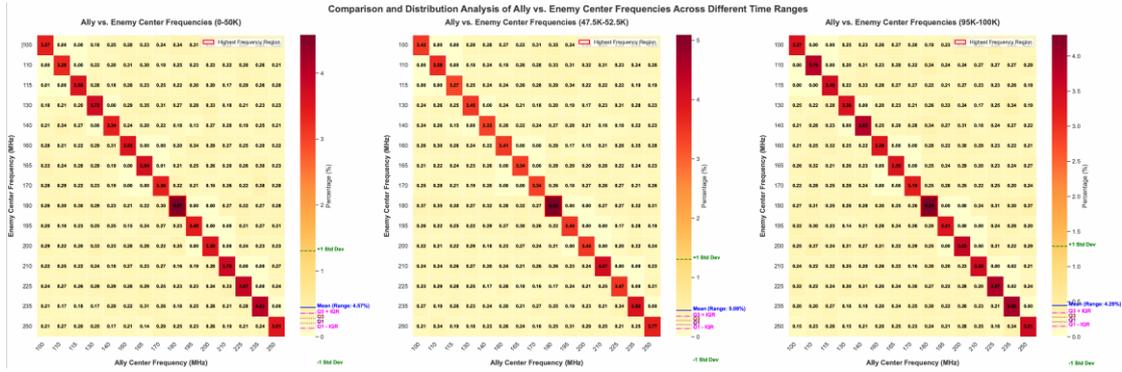

**Fig.4 Comparison of communication center frequencies between ally and opponent UAV over different time intervals.** The figure presents frequency distributions across three time segments (0-50K, 47.5K-52.5K, 95K-100K). Figure highlights the highest frequency regions (red rectangles), Q1, Q3, IQR boundaries, and Std Dev.

Simultaneously, Fig.5 illustrates the dynamic evolution of the frequency overlap ratio and jamming success rate of opponent drones over time. The figure reveals that, although the frequency overlap between the adversary and our communication center gradually decreases, the jamming success rate exhibits a continuous upward trend. This observation suggests that opponent drones may employ strategies such as utilizing broader frequency bands or comb-like spectra to interfere with the communication of our drones. Consequently, they persist in countering our communication strategies throughout the game dynamics and maintain a relatively stable reward value. This further elucidates the evolutionary characteristics of the opponent drones' jamming strategies in contested communication environments.

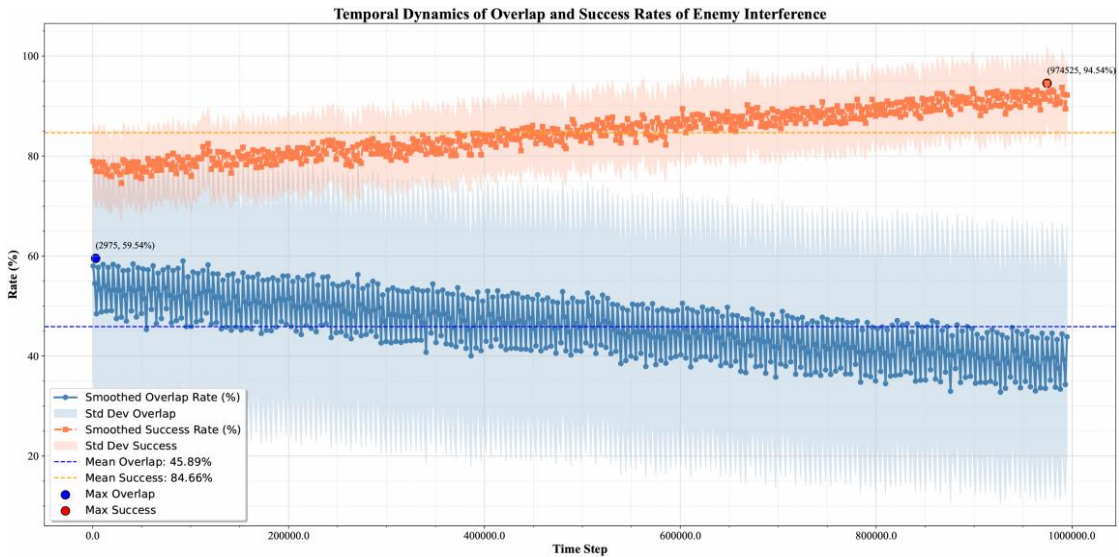

**Fig.5 Temporal evolution of opponent UAV interference frequency overlap and success rate in frequency domain competition.** A moving average smoothing procedure was applied in conjunction with sparse sampling, while standard deviation bars were included to quantify the inherent variability and uncertainty in both the overlap and success rates.

5.4 Ablation Study

In the ablation study, we divide our investigation into three main components. The first component focuses on the ablation of the opponent interference types, where a single interference method is employed to disrupt the communication between Ally UAV and the base station. This component aims to evaluate the independent effects of various interference strategies. The second component involves the ablation of the expert knowledge base, where Ally UAV operates without reliance on expert knowledge, engaging directly in a game against the adversary UAV. This analysis intends to elucidate the contribution of expert knowledge to strategic decision-making. The third component pertains to the ablation of the large language model, where reinforcement learning algorithms are used for UAV path planning, enabling the adversary UAV to navigate based on the movement directions learned through reinforcement learning. This approach allows us to assess the contribution of the large language model to the overall task performance, particularly in enhancing the ability of the adversary UAV to approach and interfere with Ally UAV.

Fig.6 displays the results of our ablation experiments on the interference methods employ by opponent UAVs. The findings reveal that the singular interference strategies utilize by the adversary generally underperform compare to Ally UAVs. Notably, the single-tone interference has the least impact, whereas the broadband jamming interference exerts the most significant effect. In the single-tone jamming scenario depicted in Fig.6a, it is observable that the allies' rewards fall below those of the adversaries during certain periods. This phenomenon can be attributed to the relatively high jamming coefficient, denoted as (k), associated with single-tone interference. Upon successful detection of the allies' communication center frequency and subsequent application of single-tone jamming by the enemy drones, significant disruption occurs in the normal communication among the allied drones, consequently leading to a reduction in their acquired rewards.

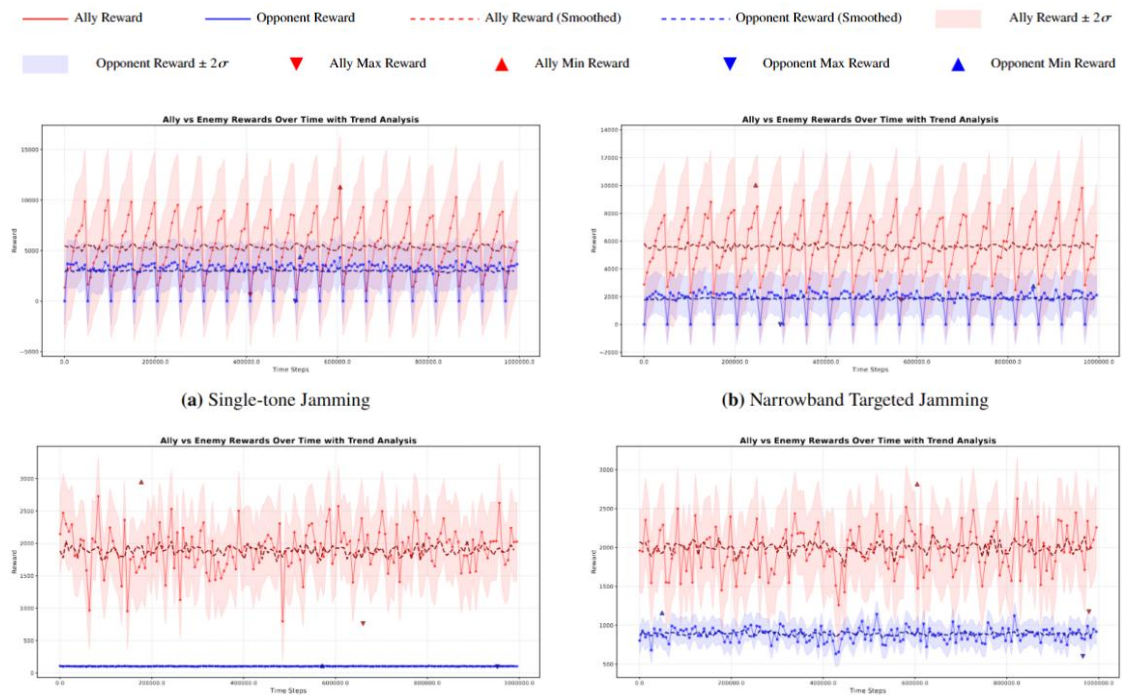

**Fig. 6 Ablation experiment on interference methods against opponent drones.** he red and blue curves represent the trends of ally and adversary rewards, respectively, with the smoothed curves (dashed lines) further capturing

the variations in reward trends. The shaded areas in the figure indicate the standard deviation range (±2σ) of the reward values.

Fig.7a and Fig.7b respectively present the ablation study results of the LLM for opponent UAV path planning and the expert knowledge base for ally UAV anti-interference strategies. The experimental results indicate that both the application of the LLM in path planning and the deployment of the expert knowledge base in anti-interference strategies play a critical role in the game process. Ablating the LLM significantly reduces the flexibility and adaptability of the opponent UAV's path, whereas ablating the expert knowledge base markedly weakens the ally UAV's anti-interference capability. These findings demonstrate that combining the LLM with the expert knowledge base effectively enhances the UAVs' performance in complex environments, improving the intelligence and adaptability of opponent encounters. Therefore, future research should focus on further optimizing the reasoning capabilities of LLMs and expanding the expert knowledge base's interference response strategies to achieve a more robust UAV opponent system.

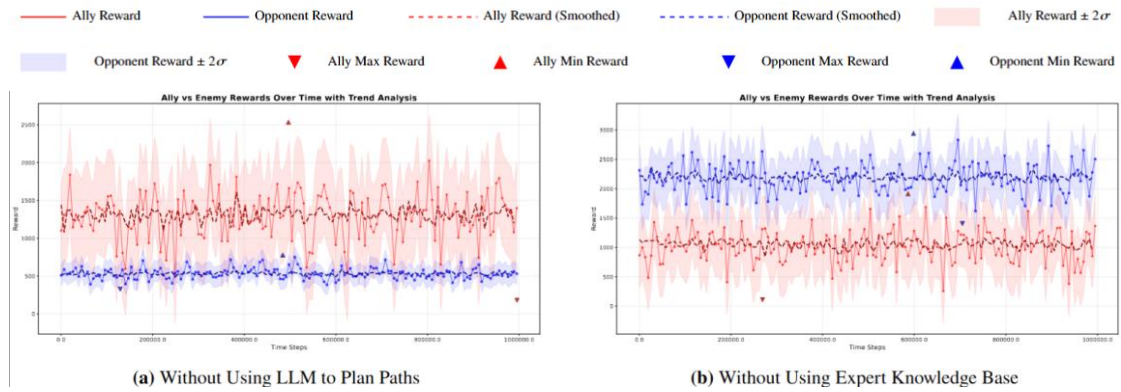

**Fig 7 Ablation Study on the Contribution of LLM-Based Path Planning and Expert Knowledge Anti-Interference.** The red and blue curves represent the rewards obtained by the ally and opponent sides in the game respectively.

## 6. Conclusion

This paper introduces **UAV-FPG**, a drone communication environment based on frequency-point game theory that is designed to optimize frequency decision-making and path planning in complex electromagnetic settings. By establishing a signal game model, we investigate the dynamic interplay between interference and anti-interference strategies of ally and opponent drones within communication frequency bands. Leveraging a reinforcement learning-based game environment, we simulate various opponent signal interference strategies alongside our own frequency hopping and spread-spectrum interference avoidance techniques, which provides an effective platform to emulate threats encountered during signal transmission. Furthermore, we integrate an expert knowledge base to enhance the decision-making capabilities of Ally UAVs in frequency selection and management, enabling more effective countermeasures against opponent interference strategies. By introducing LLMs, we augment the reasoning and signal interference abilities of opponent drones, particularly by simulating a "strong adversary" effect in path planning, thereby validating the efficacy of LLMs in dynamic path planning.

Experimental results demonstrate that the UAV-FPG environment significantly enhances the adaptability and robustness of UAVs under hostile signal interference. Specifically, in frequency-point selection experiments, as the game progresses, the overlap between Ally UAV's central frequency and the opponent's interference frequency markedly decreases, which evidences a progressive improvement in anti-interference capabilities during complex confrontations. Additionally, through multiple rounds of interaction with LLMs, we verify the substantial role of large language models in path planning; by integrating historical positions and reward information, they effectively strategize the next moves of opponent UAVs, substantially increasing the efficiency of opponent interference. Finally, through ablation studies, we further validate the independent contributions of different interference methods, expert knowledge bases, and LLMs within the UAV opponent environment. The findings indicate that expert knowledge and LLMs play pivotal roles in enhancing decision intelligence and adaptability in path planning.